\documentstyle[11pt]{article}
\begin{document}
\baselineskip=5pt
\title{\vspace{-2cm}
\bf The Role of Elliptic Operators in the Initial-Value Problem
for General Relativity}
\author{Giampiero Esposito and Cosimo Stornaiolo}
\maketitle
\hspace{-6mm}$^{1}${\em INFN, Sezione di Napoli, Complesso 
Universitario di Monte S. Angelo, Via Cintia, Edificio N',
80126 Napoli, Italy}\\
$^{2}${\em Dipartimento di Scienze Fisiche, Complesso Universitario
di Monte S. Angelo, Via Cintia, Edificio N', 80126 Napoli, Italy}

{\bf Abstract}. The
Arnowitt--Deser--Misner (ADM) equations are deeply intertwined
with discrete spectral resolutions of an elliptic operator of Laplace
type associated with the spacelike hypersurfaces which foliate the
space-time manifold, and the non-linearities of the four-dimensional
hyperbolic theory are mapped into the potential term occurring in this
operator. The ADM equations are here re-expressed as a coupled 
first-order system for the induced metric and the trace-free part
of the extrinsic-curvature tensor, and their formulation in terms
of integral equations is studied.
\vspace{1cm}

The canonical formulation of general relativity relies on the
assumption that space-time $(M,g)$ is topologically
$\Sigma \times {\bf R}$ and can be foliated by a family
of spacelike hypersurfaces $\Sigma_{t}$, all diffeomorphic to the
three-manifold $\Sigma$. The space-time metric $g$ is then 
locally cast in the form
\begin{equation}
g=-(N^{2}-N_{i}N^{i})dt \otimes dt+N_{i}(dx^{i}\otimes dt
+dt \otimes dx^{i})+h_{ij}dx^{i}\otimes dx^{j},
\label{(1)}
\end{equation}
where $N$ is the lapse function and $N^{i}$ are 
components of the shift vector of the foliation [1--3].
The induced metric $h_{ij}$ on $\Sigma_{t}$
and the associated extrinsic-curvature tensor $K_{ij}$ turn out to 
obey the first-order equations [1--3]
\begin{equation}
{\partial h_{ij}\over \partial t}=-2N K_{ij}+N_{i \mid j}
+N_{j \mid i},
\label{(2)}
\end{equation}
\begin{eqnarray} 
\; & \; & {\partial K_{ij}\over \partial t}=-N_{\mid ij}
+N \left[{ }^{(3)}R_{ij}-{ }^{(4)}R_{ij}
+K_{ij}({\rm tr}K)-2K_{im}K_{j}^{\; m}
\right] \nonumber \\
&+& \left[N^{m}K_{ij \mid m}+N_{\; \mid i}^{m}K_{jm}
+N_{\; \mid j}^{m}K_{im}\right],
\label{(3)}
\end{eqnarray}
where the stroke denotes covariant differentiation with respect
to the induced connection on $\Sigma_{t}$. Moreover, the constraint
equations hold. For Einstein theory in vacuum they read
\begin{equation}
K_{il}^{\; \; \; \mid i}-({\rm tr}K)_{\mid l} \approx 0,
\label{(4)}
\end{equation}
\begin{equation}
{ }^{(3)}R-K_{ij}K^{ij}+({\rm tr}K)^{2} \approx 0,
\label{(5)}
\end{equation}
where $\approx$ is the weak-equality symbol introduced by Dirac
to denote equations which only hold on the constraint surface [3,4].
From now on, we shall impose the vacuum Einstein equations in four
dimensions, so that in the following equations ${ }^{(4)}R_{ij}$
can be set to zero. 

Equation (3) is quasi-linear in that it contains terms quadratic
in the extrinsic-curvature tensor, i.e.
$N \left(K_{ij}({\rm tr}K)-2K_{im}K_{j}^{\; m}\right)$.
We are now aiming to obtain from eqs. (2) and (3) another set
of quasi-linear first-order equations. For this purpose, we contract
Eq. (2) with $K^{ij}$ and Eq. (3) with $h^{ij}$. This leads to
(hereafter $-\bigtriangleup \equiv -{ }_{\mid i}^{\; \; \; \mid i}$
is the Laplacian on $\Sigma_{t}$)
\begin{equation}
K^{ij}{\partial h_{ij}\over \partial t}=-2N K_{ij}K^{ij}
+2K^{ij}N_{i \mid j},
\label{(6)}
\end{equation}
\begin{eqnarray} 
\; & \; & h^{ij}{\partial K_{ij}\over \partial t}=-\bigtriangleup N
+N \left[{ }^{(3)}R +({\rm tr}K)^{2}-2K_{im}K^{im}\right] \nonumber\\
&+& \left[N^{m}({\rm tr}K)_{\mid m}+2N_{\mid i}^{m} K_{m}^{i}
\right].
\label{(7)}
\end{eqnarray}
It is now possible to obtain an evolution equation for 
$({\rm tr}K)$, because
\begin{equation}
{\partial \over \partial t}({\rm tr}K)
={\partial h^{ij}\over \partial t}K_{ij}
+h^{ij}{\partial K_{ij}\over \partial t}.
\label{(8)}
\end{equation}
The second term on the right-hand side of Eq. (8) is given
by Eq. (7), whereas the first term is obtained after using
the identity
\begin{equation}
{\partial \over \partial t}(h^{ij}h_{jl})
={\partial \over \partial t}\delta_{\; l}^{i}=0,
\label{(9)}
\end{equation}
which implies
\begin{equation}
{\partial h^{ij}\over \partial t}=-h^{ip}h^{jl}
{\partial h_{pl}\over \partial t},
\label{(10)}
\end{equation}
and hence
\begin{equation}
{\partial h^{ij}\over \partial t}K_{ij}
=-{\partial h_{ij}\over \partial t}K^{ij}.
\label{(11)}
\end{equation}
By virtue of Eqs. (6)--(8) and (11), and imposing the constraint
equation (5), we get
\begin{equation}
{\partial \over \partial t}({\rm tr}K)=-\left(\bigtriangleup
-K_{ij}K^{ij}\right)N+N^{m}({\rm tr}K)_{\mid m},
\label{(12)}
\end{equation}
which can also be cast in the form
(here ${\widehat \nabla}_{m} \equiv { }_{\mid m}$)
\begin{equation}
\left({\partial \over \partial t}-N^{m}{\widehat \nabla}_{m}\right)
({\rm tr}K)=\left(-\bigtriangleup + K_{ij}K^{ij}\right)N.
\label{(13)}
\end{equation}
Note that, without imposing the vacuum Einstein equations and the
constraint (5), the right-hand side of Eq. (13) would read instead
$$
\left(-\bigtriangleup + { }^{(3)}R+({\rm tr}K)^{2}
-{ }^{(4)}R_{ij}h^{ij}\right)N.
$$
Equation (13) is a first non-trivial result because it tells us that,
given the operator of Laplace type [5]
\begin{equation}
{\cal P} \equiv -\bigtriangleup + K_{ij}K^{ij},
\label{(14)}
\end{equation}
the right-hand side of Eq. (13) is determined by a
discrete spectral resolution of ${\cal P}$. 
By this one means a complete
orthonormal set of eigenfunctions $f_{\lambda}^{p}$ belonging
to the eigenvalue $\lambda$, so that, {\it for each fixed value} of $t$, 
the lapse can be expanded in the form
\begin{equation}
N({\vec x},t)=\sum_{\lambda}C_{\lambda}
f_{\lambda}^{p}({\vec x},t),
\label{(15)}
\end{equation}
with Fourier coefficients $C_{\lambda}$ given by the scalar
product $C_{\lambda}=\Bigr(f_{\lambda}^{p},N \Bigr)$.
This point is simple but non-trivial: for each fixed value of $t$,
the restriction of the lapse to $\Sigma_{t}$ becomes a function
on $\Sigma_{t}$ only, and hence the Fourier coefficients in (15) read
\begin{equation}
C_{\lambda}=\int_{\Sigma_{t}}(f_{\lambda}^{p})^{*}N \sqrt{h}d^{3}x
\equiv C_{\lambda,t},
\label{(16)}
\end{equation}
where the star denotes complex conjugation. In the application
to the initial-value problem, one studies the lapse at different
values of $t$, and hence it is better to write its expansion in
the form (15), with the time parameter explicitly included.
The integration measure for $C_{\lambda}$, however, remains
the invariant integration measure on $\Sigma_{t}$.

The existence of discrete spectral resolutions 
of ${\cal P}$ is guaranteed
if $\Sigma_{t}$ is a compact Riemannian manifold without 
boundary [5], which is what we assume hereafter. Thus, Eq. (13)
can be re-expressed in the form
\begin{equation}
\left({\partial \over \partial t}-N^{m}{\widehat \nabla}_{m}
\right){\rm tr}K({\vec x},t)
=\sum_{\lambda}\lambda C_{\lambda,t}
f_{\lambda}^{p}({\vec x},t).
\label{(17)}
\end{equation}
On denoting by $L$ the operator
\begin{equation}
L \equiv {\partial \over \partial t}
-N^{m}{\widehat \nabla}_{m},
\label{(18)}
\end{equation}
and writing $G({\vec x},t;{\vec y},t')$
for the Green function of $L$, Eq. (17) can
be therefore solved for $({\rm tr}K)$ in the form
\begin{equation}
{\rm tr}K({\vec x},t)
=\int_{\Sigma_{t'}\times {\bf R}} G({\vec x},t;{\vec y},t')
\sum_{\lambda}\lambda C_{\lambda,t'}f_{\lambda}^{p}({\vec y},t')
d\mu({\vec y},t'),
\label{(19)}
\end{equation}
where $d\mu({\vec y},t')$ is the invariant integration measure
on $\Sigma_{t'} \times {\bf R}$. Note that the eigenfunctions
$f_{\lambda}^{p}$ are different at different times and hence,
to compute them at all times, one has to integrate the equations
of motion so that the metric and the extrinsic-curvature tensor
are known at all times. Thus, Eq. (19) does not reduce the amount
of entanglement of the original equations (2) and (3).
A relevant particular case is obtained on considering the canonical
form of the foliation [6], for which the shift vector vanishes. The
operator $L$ reduces then to ${\partial \over \partial t}$. 

A cosmological example shows how the right-hand side of Eq. (17) can
be worked out explicitly in some cases. We are here concerned with
closed Friedmann--Lemaitre--Robertson--Walker models, for which the
three-manifold $\Sigma$ reduces to a three-sphere of radius $a$. 
It is indeed well known that the space of functions on the 
three-sphere can be decomposed by using the invariant subspaces 
corresponding to irreducible representations of $O(4)$, the
orthogonal group in four dimensions. These 
invariant subspaces are spanned by 
the hyperspherical harmonics [7], i.e.
generalizations to the three-sphere of the familiar spherical 
harmonics. The scalar hyperspherical harmonics 
$Q^{(n)}(\chi,\theta,\varphi)$ form a basis spanning the invariant
subspace labeled by the integer $n \geq 1$ corresponding to the
$n$-th scalar representation of $O(4)$. The label $(n)$ of $Q^{(n)}$
refers both to the order $n$ and to other labels denoting the 
different elements spanning the subspace. The number of elements 
$Q^{(n)}$ is determined by the dimension of the corresponding 
$O(4)$ representation. On using the general formula for $O(2k)$
representations, the dimension of the irreducible scalar 
representations is found to be $d_{Q}(n)=n^{2}$. The Laplacian
on a unit three-sphere when acting on $Q^{(n)}$ gives
\begin{equation}
-\bigtriangleup Q^{(n)}=(n^{2}-1)Q^{(n)} \; \; \;
n=1,2,... \; \; .
\label{(20)}
\end{equation}
Any arbitrary function $F$ on the three-sphere can be expanded in
terms of these hyperspherical harmonics as they form a complete set:
\begin{equation}
F=\sum_{(n)}q_{(n)}Q^{(n)},
\label{(21)}
\end{equation}
where the sum $(n)$ runs over $n=1$ to $\infty$ and over the $n^{2}$
elements in each invariant subspace, and the $q_{(n)}$ are constant
coefficients. The lapse function on the three-sphere 
is therefore expanded according to (cf. (15))
\begin{equation}
N=\sum_{(n)}c_{(n)}Q^{(n)},
\label{(22)}
\end{equation}
and bearing in mind that $K_{ij}K^{ij}=3$ on a unit three-sphere
we find, on a three-sphere of radius $a$,
\begin{equation}
{\cal P}N=\sum_{(n)}c_{(n)}\lambda_{n}Q^{(n)},
\label{(23)}
\end{equation}
where (here $a=a(t)$)
\begin{equation}
\lambda_{n}={(n^{2}+2)\over a^{2}} \; \; n=1,2,... \; \; .
\label{(24)}
\end{equation}
We have therefore found that all nonlinearities of the original 
equations give rise to the potential term in the operator ${\cal P}$
defined in Eq. (14), which is a positive-definite operator of 
Laplace type (unlike the Laplacian, which is bounded from below
but has also a zero eigenvalue when $n=1$).

The interest in the evolution equation for the trace of the 
extrinsic-curvature tensor is motivated by a careful analysis of the
variational problem in general relativity. More precisely, the
``cosmological action'' is sometimes considered, which is the
form of the action functional $I$ appropriate for the case when
$({\rm tr}K)$ and the conformal three-metric ${\widetilde h}_{ij}
\equiv h^{-{1\over 3}} \; h_{ij}$ 
are fixed on the boundary. One then finds in $c=1$ units [8]
\begin{equation}
I={1\over 16 \pi G}\int_{M}{ }^{(4)}R\sqrt{-g} \; d^{4}x
+{1\over 24 \pi G}\int_{\partial M}({\rm tr}K)\sqrt{h} \; d^{3}x,
\label{(25)}
\end{equation}
and this formula finds important applications also to the quantum
cosmology of a closed universe, giving rise to the
$K$-representation of the wave function of the universe [9].

The program we have outlined in our paper shows an intriguing link 
between the Cauchy problem in general relativity  
on the one hand [10--13], and the spectral theory of elliptic
operators of Laplace type on the other hand [5], which is
obtained by exploiting the nonlinear form of the ADM equations
and the nonpolynomial form of the Hamiltonian constraint (5). 
On denoting by $\sigma_{ij}$ the trace-free part of the 
extrinsic-curvature tensor:
\begin{equation}
\sigma_{ij} \equiv K_{ij}-{1\over 3}h_{ij}({\rm tr}K)
\label{(26)}
\end{equation}
we therefore obtain a set of equations, equivalent to (2) and (3),
in the form (here $L^{-1}$ denotes the inverse of the operator
$L$ defined in (18))
\begin{equation}
{\rm tr}K=L^{-1} \; {\cal P}N,
\label{(27)}
\end{equation}
\begin{equation}
{\partial h_{ij}\over \partial t}=-{2\over 3}Nh_{ij}({\rm tr}K)
-2N \sigma_{ij}+2N_{(i \mid j)},
\label{(28)}
\end{equation}
\begin{equation}
{\partial \sigma_{ij}\over \partial t}=
\Omega_{ij}N-{2\over 3}N_{(i \mid j)}({\rm tr}K)
+N^{m}{\widehat \nabla}_{m}\sigma_{ij} 
+2N_{(\mid i}^{m}\Bigr[\sigma_{j)m}+{1\over 3}h_{j)m}
({\rm tr}K)\Bigr], 
\label{(29)} 
\end{equation}
where the operator of Laplace type $\Omega_{ij}$ is given by
\begin{eqnarray} 
\Omega_{ij}& \equiv & -{\widehat \nabla}_{j}{\widehat \nabla}_{i}
-{1\over 3}h_{ij}{\cal P}+{1\over 3}\sigma_{ij}({\rm tr}K) \nonumber\\
&+& {1\over 3}h_{ij}({\rm tr}K)^{2}+{ }^{(3)}R_{ij}
-2 \sigma_{im}\sigma_{j}^{\; m}.
\label{(30)}
\end{eqnarray}
Hereafter, to be more rigorous, we assume that the left and right
inverses of $L$ exist and are the same, and that left and right inverses
coincide for all operators we have to consider.

It now appears desirable to understand to which extent the
differential-operator viewpoint plays a role in solving the coupled
system (27)--(30). For this purpose, we consider a relevant
particular case, i.e. the canonical foliation studied in Ref. [6],
for which the shift vector vanishes. The operator $L$ reduces
then to ${\partial \over \partial t}$ and, on defining 
the operators
\begin{equation}
Q \equiv {\partial \over \partial t}+{2\over 3}N{\rm tr}K 
=L+{2\over 3}NL^{-1}{\cal P}N,
\label{(31)}
\end{equation}
\begin{equation}
S \equiv {\partial \over \partial t}-{N\over 3}{\rm tr}K
=L-{N\over 3}L^{-1}{\cal P}N,
\label{(32)}
\end{equation}
\begin{equation}
A_{ij} \equiv -{\widehat \nabla}_{j}{\widehat \nabla}_{i}
+{ }^{(3)}R_{ij},
\label{(33)}
\end{equation}
one finds the equations
\begin{equation}
\left[S+{2\over 3}(L^{-1}{\cal P}N)^{2}Q^{-1}N \right]\sigma_{ij}
-{2\over 3}(Q^{-1}N \sigma_{ij}){\cal P}N 
+2N \sigma_{im}\sigma_{j}^{\; m}=A_{ij}N,
\label{(34)} 
\end{equation}
\begin{equation}
h_{ij}=-2Q^{-1}N \sigma_{ij},
\label{(35)}
\end{equation}
supplemented, of course, by Eq. (27). In other words, once Eq. (34) is
solved for the trace-free part of the extrinsic-curvature tensor of
$\Sigma_{t}$, the induced metric on $\Sigma_{t}$ is obtained from 
Eq. (35). This form of the coupled first-order ADM system of equations
has possibly the merit of stressing the need to invert the operators
$L,Q$ and $S$ to find a solution for given initial conditions. 
Approximate solutions, to the desired accuracy, will correspond to
the construction of approximate inverse operators $L^{-1},Q^{-1}$ and
$S^{-1}$. For this purpose, it can be useful to re-express Eq. (34)
in the form ($I$ being the identity operator)
\begin{eqnarray}
\; & \; & \left[I+{2\over 3}S^{-1}(L^{-1}{\cal P}N)^{2}Q^{-1}N \right]
\sigma_{ij}-{2\over 3}S^{-1}\left[(Q^{-1}N \sigma_{ij})
{\cal P}N \right] \nonumber\\
&-& S^{-1}(A_{ij}N)=-2S^{-1}\left(N \sigma_{im}\sigma_{j}^{\; m}\right).
\label{(36)}
\end{eqnarray}
The action of the inverse operator $L^{-1}$ on ${\cal P}N$ is already 
given by Eq. (19), but the inverses $Q^{-1}$ and $S^{-1}$ make it
necessary to develop an algorithm for their exact or approximate
evaluation. More precisely, one has
\begin{equation}
Q=L \left(I+{2\over 3}L^{-1}NL^{-1}{\cal P}N \right),
\label{(37)}
\end{equation}
\begin{equation}
S=L \left(I-{1\over 3}L^{-1}NL^{-1}{\cal P}N \right),
\label{(38)}
\end{equation}
and hence
\begin{equation}
Q^{-1}=\left(I+{2\over 3}L^{-1}NL^{-1}{\cal P}N \right)^{-1}L^{-1},
\label{(39)}
\end{equation}
\begin{equation}
S^{-1}=\left(I-{1\over 3}L^{-1}NL^{-1}{\cal P}N \right)^{-1}L^{-1},
\label{(40)}
\end{equation}
by virtue of the operator identity $(AB)^{-1}=B^{-1}A^{-1}$. From this
point of view, two levels of approximation seem to emerge:
\vskip 0.3cm
\noindent
(i) The degree of accuracy in the evaluation of $Q^{-1}$ and $S^{-1}$.
It is clear from Eqs. (34) and (35) that this reduces to finding 
$\left(I+\rho L^{-1}N L^{-1}{\cal P}N \right)^{-1}$,
where $\rho={2\over 3}$ or $-{1\over 3}$. A series expansion will be,
in general, only of formal value. However, on using the norms defined
in section 7.4 of Ref. [14], which relies on well known properties
of Sobolev spaces, if the operator
\begin{equation}
T_{\rho} \equiv \rho L^{-1}NL^{-1}{\cal P}N
\label{(41)}
\end{equation}
is a bounded operator on a Banach space $X$ with norm
$\left \| T_{\rho} \right \| < 1$, the inverse of $I+T \rho$ exists
and the series $\sum_{n=0}^{\infty}(-1)^{n}(T_{\rho})^{n}$ converges
uniformly to $(I+T_{\rho})^{-1}$ with respect to the norm on the 
set of bounded maps from $X$ into $X$, and one can write
\begin{equation}
(I+T_{\rho})^{-1}=\sum_{n=0}^{\infty}(-1)^{n}(T_{\rho})^{n}.
\label{(42)}
\end{equation}
Equation (42) can be therefore used if $L^{-1}$ 
and $\cal P$ are such that 
$T_{\rho}$ is a bounded operator with 
$\left \| T_{\rho} \right \| < 1$.
\vskip 0.3cm
\noindent
(ii) The way in which the non-linear term $\sigma_{im}\sigma_{j}^{\; m}$
is dealt with. For example, one may regard the right-hand side of Eq.
(36) as the ``known term'', and hence consider the equation
\begin{eqnarray}
\; & \; & \sigma_{ij}dx^{i} \otimes dx^{j}-{2\over 3}F^{-1}S^{-1}
\left[(Q^{-1}N \sigma_{ij}){\cal P}N \right]dx^{i} \otimes dx^{j} \nonumber\\
&-& F^{-1}S^{-1}(A_{ij}N)dx^{i} \otimes dx^{j}
=-2F^{-1}S^{-1}(N \sigma_{im}\sigma_{j}^{\; m})
dx^{i} \otimes dx^{j},
\label{(43)}
\end{eqnarray}
where $F^{-1}$ is the inverse of the operator
\begin{equation}
F \equiv I+{2\over 3}S^{-1}(L^{-1}{\cal P}N)^{2}Q^{-1}N.
\label{(44)}
\end{equation}
Equation (43) is an integral equation for $\sigma_{ij}$, for which a
perturbation approach might be useful if one takes
$-2F^{-1}S^{-1}\left(N \sigma_{im}\sigma_{j}^{\; m}\right)
dx^{i} \otimes dx^{j}$
as the known term mentioned before. Note that we have resorted to the
use of the tensor product $dx^{i} \otimes dx^{j}$ because in the
resulting equation (43) one has a well defined integration of a
symmetric rank-two tensor field over the space-time manifold, here taken to
be diffeomorphic to $\Sigma_{t} \times {\bf R}$. Strictly, also
Eq. (35) should be written as
\begin{equation}
h_{ij}dx^{i} \otimes dx^{j}=-2 \Bigr(Q^{-1}N \sigma_{ij}\Bigr)
dx^{i} \otimes dx^{j},
\label{(45)}
\end{equation}
and in all such equations only the symmetric part of the tensor product
survives, because both $h_{ij}$ and $\sigma_{ij}$ are symmetric
rank-two tensor fields.

Note once more that it would be wrong
to regard Eq. (27) as independent of the solution of Eqs. (28)
and (29), because $L^{-1} \; {\cal P}N$ is only known at all times
when the time evolution of $h_{ij}$ and $K_{ij}$ has been determined
for given initial conditions. 
Our equations (43) and (45), 
although not obviously more powerful than
previous schemes, prepare the ground for an operator approach to the
ADM equations, and hence might contribute
to the understanding of structural properties of general relativity.
In particular, our way of writing the coupled system
of non-linear evolution equations might lead to
a better understanding of the interplay between elliptic
operators on the spacelike hypersurfaces $\Sigma_{t}$ and hyperbolic
equations on the space-time manifold. As far as we can see, this
expectation is supported by the closed 
Friedmann--Lemaitre--Robertson--Walker model discussed before
(where $\sigma_{ij}$ vanishes), 
and further examples of cosmological interest might be found,
e.g. Bianchi IX models describing a closed but anisotropic universe.
For open universes or asymptotically flat space-times, however,
we are not aware of theorems that make it possible to expand the
lapse as in Eq. (15). In such cases, the counterpart
of the elliptic theory on
$\Sigma_{t}$ advocated in our paper is therefore another
open problem [15].
\vskip 0.3cm
\noindent
$\; \;$ {\bf Acknowledgments}. The present paper is dedicated to
the loving memory of Ruggiero de Ritis, who introduced us to
General Relativity when we were young students and encouraged our
research with great enthusiasm.

\vspace{1cm}
\begin{description}

\item [{[1]}]
R. Arnowitt, S. Deser and C. W. Misner, in {\it Gravitation: an
Introduction to Current Research}, 
edited by L. Witten (Wiley, New York, 1962).
\item [{[2]}]
C. W. Misner, K. P. Thorne and J. A. Wheeler, {\it Gravitation}
(Freeman, S. Francisco, 1973).
\item [{[3]}]
G. Esposito, {\it Quantum Gravity, Quantum Cosmology and 
Lorentzian Geometries}, Lecture Notes in Physics, Vol. m12,
Second Corrected and Enlarged Edition (Springer--Verlag,
Berlin, 1994).
\item [{[4]}]
P. A. M. Dirac, {\it Lectures on Quantum Mechanics}, Belfer Graduate
School (Yeshiva University, New York, 1964).
\item [{[5]}]
P. B. Gilkey, {\it Invariance Theory, the Heat Equation and
the Atiyah--Singer Index Theorem} (CRC Press, Boca Raton, 1995).
\item [{[6]}]
D. Christodoulou and S. Klainerman, {\it The Global Nonlinear Stability
of the Minkowski Space} (Princeton University Press, Princeton, 1993).
\item [{[7]}] 
E. M. Lifshitz and I. M. Khalatnikov, {\it Adv. Phys.} {\bf 12},
185 (1963).
\item [{[8]}]
J. W. York, {\it Found. Phys.} {\bf 16}, 249 (1986).
\item [{[9]}]
J. B. Hartle and S. W. Hawking, {\it Phys. Rev.} {\bf D 28},
2960 (1983).
\item [{[10]}]
D. Christodoulou, {\it Class. Quantum Grav.} {\bf 16},
A23 (1999).
\item [{[11]}]
A. Anderson and J. W. York, {\it Phys. Rev. Lett.} {\bf 82},
4384 (1999).
\item [{[12]}]
A. Anderson, Y. Choquet-Bruhat and J. W. York, ``Einstein's
Equations and Equivalent Hyperbolic Dynamical Systems''
(GR-QC 9907099).
\item [{[13]}]
H. Friedrich and A. Rendall, in {\it Einstein's Field Equations
and their Physical Interpretation}, edited by B. G. Schmidt
(Springer--Verlag, Berlin, 2000; GR-QC 0002074).
\item [{[14]}]
S. W. Hawking and G. F. R. Ellis, {\it The Large-Scale Structure
of Space-Time} (Cambridge University Press, Cambridge, 1973).
\item [{[15]}]
G. Esposito and C. Stornaiolo, {\it Found. Phys. Lett.}
{\bf 13}, 279 (2000).
\end{description}
\end{document}